\def\beq{\begin{equation}}\def\eeq{\end{equation}}
\def\N{{\cal N}}
\def\xb{\bar x}
\def\s{\sigma}\def\g{\gamma}\def\l{\lambda}
\documentstyle[12pt]{article}
\begin{document}
\begin{titlepage}
\title{Tentative statistical interpretation of non-equilibrium entropy}
\author{Lajos Di\'osi\thanks{E-mail: diosi@rmki.kfki.hu}\\
Research Institute for Particle and Nuclear Physics\\
H-1525 Budapest 114, POB 49, Hungary}
\maketitle
\begin{abstract}
We suggest a certain statistical interpretation for the entropy produced in 
driven thermodynamic processes. The exponential function of \emph{half} 
irreversible entropy re-weights the probability of the standard 
Ornstein-Uhlenbeck-type thermodynamic fluctuations. (We add a proof of the 
standard Fluctuation Theorem which represents a more natural interpretation.)

\end{abstract}
\end{titlepage}

In 1910 Einstein \cite{Ein}, paraphrasing \cite{Wpm} Boltzmann's lapidary 
formula 
$S=\log W$, expressed the probability distribution of thermodynamic variables 
$x$ through the entropy function $S(x)$:
\beq\label{Ein}
W(x) \sim e^{S(x)}~.
\eeq
This equation describes thermodynamic fluctuations in Gaussian approximation 
properly. Going beyond the stationary features, the time-de\-pen\-dence of 
fluctuations $x_t$ can be characterized by a certain probability functional 
$W[x]$ over complete paths \hbox{$\{x_t;~t\in(-\infty,\infty)\}$}. It turns 
out that, in driven thermodynamic processes, this probability is related to 
the irreversible entropy $S_{irr}[x]$. Symbolically, we can write the 
following relationship:
\beq\label{Dio}
W[x] \sim W_{OU}[x-\xb] e^{S_{irr}[x]/2}~,
\eeq
where $\xb_t$ is the `driving' value of parameter $x_t$ and $W_{OU}[z]$ turns 
out to correspond to fluctuations $z_t$ of Ornstein-Uhlenbeck type. This 
relationship offers $S_{irr}$ a certain statistical interpretation, somehow 
resembling Einstein's suggestion (\ref{Ein}) for the equilibrium entropy 
$S(x)$. In this short note, Einstein's approach to the thermodynamic 
fluctuations is outlined and standard equations of time-dependent fluctuations 
are invoked from irreversible thermodynamics. Then I give a precise form to 
the relationship (\ref{Dio}) for driven thermodynamic processes.

The equilibrium conditions for isolated composite thermodynamic systems derive 
from the maximum entropy principle:   
\beq\label{Smax}
S(x)=max~,
\eeq
where $S(x)$ is the total entropy of the system in function of certain free 
thermodynamic parameters $x$ \cite{LanLif}. If the function $S(x)$ is maximum 
at $x=\xb$ then $\xb$ is the equilibrium state. For example, $x$ may be the 
temperature $T$ of a small (yet macroscopic) subsystem in the large isolated 
system of temperature $\overline{T}=\xb$. Then, the function $S(x)$ must be 
the total entropy of the isolated system, depending on the variation of the 
subsystem's temperature around its equilibrium value. The equilibrium value 
$\xb$ [as well as $S(x)$ itself] may vary with the deliberate alteration of 
the initial conditions. Surely, in our example the temperature $\overline{T}$ 
of the whole isolated system can always be controlled at will. For later 
convenience, especially in treating driven thermodynamic processes, we may 
prefer the explicit detailed notation $S(x|\xb)$ for $S(x)$. Though 
$S(\xb)-S(x)$ might qualify the lack of equilibrium, nearby values 
$x\approx\xb$ have no interpretation in phenomenological thermodynamics. They 
only have it in the broader context of statistical physics. In finite 
thermodynamic systems there are fluctuations around the equilibrium state 
$\xb$ and their probability follows Eq.~(\ref{Ein}):  
\beq\label{Einxb}
W(x|\xb)dx= \N e^{S(x|\xb)-S(\xb|\xb)}dx~.
\eeq
Assume, for simplicity, that there is a single free variable $x$. The Taylor 
expansion of the entropy function yields Gaussian fluctuations:
\beq\label{EinGau}
W(x|\xb)= 
\frac{1}{\sqrt{2\pi\s^2}} 
\exp\Bigl(-{1\over2\s^2}\bigl(x-\xb\bigl)^2\Bigl)~,
\eeq
where 
\beq\label{sigma}
\frac{1}{\s^2}=-{S^{\prime\prime}(\xb)}
\equiv-\frac{\partial^2S(x\vert\xb)}{\partial x^2}\Big\vert_{x=\xb}~. 
\eeq
In our concrete example $\s^2=T^2/C$ where $C$ is the specific heat of the 
subsystem.

We are going to regard the time-de\-pen\-dence of the parameter $x_t$ 
fluctuating around $\xb$, according to the standard irreversible 
thermodynamics \cite{LanLif}. The time-dependent fluctuation 
$z_t\equiv x_t-\xb$ is an Ornstein-Uhlenbeck (OU) stochastic process 
\cite{OrnUhl} of zero mean $\langle z_t\rangle\equiv0$ and of correlation
\beq\label{OUcorr}
\langle~z_t~z_{t^\prime}~\rangle_{OU}=\s^2
e^{-\l\vert t-t^\prime \vert}~.
\eeq
The relaxation rate $\l$ of fluctuations is related to the corresponding 
Onsager kinetic constant $\g$ by $\l=\g/\s^2$. It can be shown that the 
probability distribution of $x_t=z_t+\xb$ at any fixed time $t$ is the 
Gaussian distribution (\ref{EinGau}) as it must be. For the probability of the 
complete fluctuation path $z_t$, the zero mean and correlation (\ref{OUcorr}) 
are equivalent with the following functional: 
\beq\label{WOU}
W_{OU}[z]{\cal D}z=\exp\Bigl(
-{1\over4\g}\int~(\dot z_t^2+\l^2 z_t^2)dt\Bigr){\cal D}z~,
\eeq
where a possible constant of normalization has been absorbed into the
functional measure ${\cal D}z$.

In order to construct and justify a relationship like (\ref{Dio}) one needs to 
proceed to driven thermodynamic processes. In fact, we assume that we are 
varying the parameter $\xb$ with small but finite velocity. Formally, the 
parameter $\xb$ becomes time-dependent. For simplicity's sake we assume that 
the coefficients $\s,\g$ do not depend on $\xb$ or, at least, that we can 
ignore their variation throughout the driven range of $\xb_t$. We define the 
irreversible entropy production during the driven process as follows:
\beq\label{Sirr1}                                          
S_{irr}[x|\xb]=\frac{1}{\s^2}\int~(\xb_t-x_t)dx_t~.
\eeq
In our concrete example 
$dS_{irr}=(C/T^2)(\overline{T}-T)dT\approx dQ(T^{-1}-\overline{T}^{-1})$ which 
is indeed the entropy produced randomly by the heat transfer $dQ$ from the 
surrounding to the subsystem. By partial integration, Eq.~(\ref{Sirr1}) leads 
to an alternative form:
\beq\label{Sirrparc}
S_{irr}[x|\xb]=\frac{1}{\s^2}\int~(\xb_t-x_t)d\xb_t
              +\frac{1}{\s^2}(x_{-\infty}-\xb_{-\infty})^2              
              -\frac{1}{\s^2}(x_{\infty}-\xb_{\infty})^2~.              
\eeq
In relevant driven processes the entropy production is macroscopic, i.e., 
$S_{irr}\gg1$ in $k_B$-units, hence it is dominated by the integral term 
above. I exploit this fact to replace expression (\ref{Sirr1}) by 
\beq\label{Sirr2}
S_{irr}[x|\xb]=\frac{1}{\s^2}\int~(\xb_t-x_t)d\xb_t
\eeq
which vanishes for constant $\xb$. In the sense of the guess 
(\ref{Dio}), I suggest the following form for the probability distribution of 
the driven path:
\beq\label{Dioxb}
W[x|\xb]=
\N[\xb] W_{OU}[x-\xb]e^{S_{irr}[x|\xb]/2}~.
\eeq
The non-trivial normalizing pre-factor is a consequence of $\xb$'s 
time-de\-pen\-dence and will be derived below. Since the above distribution is 
a Gaussian functional and $S_{irr}[x|\xb]$ is a linear functional 
(\ref{Sirr2}) of $x$, we can easily calculate the expectation value of the 
irreversible entropy:
\beq\label{Sirrmean}
S_{irr}[\xb]\equiv\langle S_{irr}[x|\xb] \rangle
=\frac{1}{2\s^2}
        \int\int\dot{\xb}_t\dot{\xb}_{t^\prime}
        e^{-\l\vert t-t^\prime \vert}dtdt^\prime~.
\eeq
In case of moderate accelerations $\ddot{\xb}\ll\l\dot{\xb}$, this expression 
reduces to the standard irreversible entropy 
$\g^{-1}\!\int\dot{\xb}_t^2dt$ of the phenomenological theory of driven 
processes \cite{Dio4}. Coming back to the normalizing factor in 
Eq.~(\ref{Dioxb}), we can relate it to the mean entropy production 
(\ref{Sirrmean}): \mbox{$\N[\xb]=exp(-S_{irr}[\xb]/4)$}. 
Hence, the ultimate form of Eq.~(\ref{Dioxb}) will be:
\beq\label{Dioresult}
W[x|\xb]=
W_{OU}[x-\xb]e^{S_{irr}[x|\xb]/2-S_{irr}[\xb]/4}~.
\eeq

This result gives the precise meaning to our symbolic relationship 
(\ref{Dio}). If the entropy production $S_{irr}$ were negligible then the
thermodynamic fluctuations $x_t-\xb_t$ would follow the OU statistics 
(\ref{OUcorr}) like in case of a steady state $\xb_t=const$. Even in slow 
irreversibly driven processes $S_{irr}$ may grow essential and 
$exp[S_{irr}/2]$ 
will re-weight the probability of OU fluctuations. The true stochastic 
expectation value of an arbitrary functional $F[x]$ can be expressed by the OU 
expectation values of the re-weighted functional:
\beq\label{Fmean}
\left\langle F[x]\right\rangle 
=\Bigl\langle 
F[x]e^{S_{irr}[x|\xb]/2-S_{irr}[\xb]/4}
\Bigr\rangle_{OU}~.
\eeq

I can verify the plausibility of Eq.~(\ref{Dioresult}) for the special case of 
small accelerations. Let us insert Eqs.~(\ref{WOU},\ref{Sirr2}) and also 
Eq.~(\ref{Sirrmean}) while ignore $\ddot{\xb}$ in comparison with 
$\l\dot{\xb}$. We obtain:
\beq\label{WOUret}
W[x|\xb]=
W_{OU}[x-\xb+\l^{-1}\dot{\xb}]~.
\eeq
Obviously, the fluctuations of the driven system are governed by the OU 
process $z_t$ (\ref{OUcorr}) in the equilibrium case when $\dot{\xb}\equiv0$. 
In driven process, when $\dot{\xb}\neq0$, there is only a simple change: 
The OU fluctuations happen around the retarded value
$\xb_t-\tau\dot{\xb}\approx\xb_{t-\tau}$ of the driven parameter. The lag 
$\tau~$ is equal to the thermodynamic relaxation time $1/\l$. Consequently, 
the driven random path takes the following form:
\beq\label{xtret}
x_t=\xb_{t-\tau}+z_t~,
\eeq
where $z_t$ is the equilibrium OU process (\ref{OUcorr}). This result implies, 
in particular, the equation $\langle x_t\rangle=\xb_{t-\tau}$ which is just 
the retardation effect well-known in the thermodynamic theory of slightly 
irreversible driven processes. For example, in case of an irreversible heating 
process the subsystem's average temperature will always be retarded by 
$\tau~\dot{\overline{T}}$ with respect to the controlling temperature 
$\overline{T}$ \cite{Dio4}.

Finally, let us summarize the basic features of Einstein's formula (\ref{Ein})
and of the present proposal (\ref{Dio}). They characterize the quality of 
equilibrium in static and in driven steady states, respectively.
They do it in terms of thermodynamic entropies while they refer to a 
statistical context lying outside both reversible and irreversible 
thermodynamics. Both formulae are only valid in the lowest non-trivial order 
and their correctness in higher orders is questionable \cite{Langevin}. 
Contrary to their limited validity, they can no doubt give an insight into the 
role of thermodynamic entropy in statistical fluctuations around both 
equilibrium or non-equilibrium states. 

{\it Acknowledgments.\/}
I thank Bjarne Andresen, Karl Heinz Hoffmann, Attila R\'acz, and Stan 
Sieniutycz  for useful remarks regarding the problem in general. This work 
enjoyed the support of the EC Inco-Copernicus program {\it Carnet 2}. 

{\it Note added.\/}
The first version of this work proposed the relationship
$$
W\sim W_{OU}e^{S_{irr}}~,
$$
the exponent was free from the funny factor $1/2$. The proof was wrong, 
of course. With the factor $1/2$, my statistical interpretation for $S_{irr}$ 
has become less attractive. I also realized that a more natural statistical 
interpretation \cite{FT} was already discovered before. The present
formalism offers the following convenient proof of the Fluctuation Theorem. 

The true probability distribution of the slowly driven process is the 
Onsager--Machlup functional \cite{OnsMac53}:
$$
W_{OM}[x\vert\xb]{\cal D}x=\exp\Bigl(
-{1\over4\g}\int~[\dot x_t+\l(x_t-\xb_t)]^2dt\Bigr){\cal D}x~,
$$
at fixed $x_{-\infty}$.
Let us imagine the probability distribution of the time-reversed process 
$x_t^r=x_{-t}$ driven by the time-reversed surrounding 
$\xb_t^r=\xb_{-t}$. Formally, we only have to change the sign of 
$\dot x_t$, yielding:
$$ 
W_{OM}[x^r\vert\xb^r]{\cal D}x=\exp\Bigl(
-{1\over4\g}\int~[\dot x_t-\l(x_t-\xb_t)]^2dt\Bigr){\cal D}x~,
$$
at fixed $x^r_{-\infty}$.
We can inspect that the above distributions of the true and the 
time-reversed processes, respectively, satisfy the following relationship:
$$
\log  W_{OM}[x\vert\xb] - \log W_{OM}[x^r\vert\xb^r]
=\frac{\l}{\g}\int(\xb_t-x_t)dx_t~.
$$ 
Observe that the r.h.s. is the irreversible entropy production
$S_{irr}[x\vert\xb]$ of the driven process. This leads to the
so-called Fluctuation Theorem:
$$
W_{OM}[x^r\vert\xb^r]=e^{-S_{irr}[x\vert\xb]}W_{OM}[x\vert\xb]~.
$$
The irreversible entropy turns out to be a concrete statistical
measure of the time-reversal asymmetry.


\begin{thebibliography}{99}
%
\bibitem{Ein} A. Einstein, Ann.Phys.(Leipzig) {\bf 33}, 1275 (1910).
%
\bibitem{Wpm} The letter $W$ stood for phase volume in Boltzmann relation 
while it denotes probability in (\ref{Ein}). I am grateful to Jiri Vala 
who showed me that Einstein \cite{Ein}, maybe for somehow related reasons, 
committed (eventually innocent) sign errors repeatedly confusing 
$e^W$ and $e^{-W}$.
%
\bibitem{LanLif} L.D. Landau and E.M. Lifshitz, {\it Statistical Physics}
(Clarendon, Oxford, 1982).
%
\bibitem{OrnUhl} G.E.Uhlenbeck and L.S.Ornstein, Phys.Rev. {\bf 36}, 823 
(1930).
%
\bibitem{Dio4} L.Di\'osi, Katalin Kulacsy, B.Luk\'acs and A.R\'acz,
J.Chem.Phys. {\bf 105}, 11220 (1996).
%
\bibitem{Langevin} Einstein's ansatz fails obviously beyond the Gaussian 
approximation. Our present proposal is first of all limited to small 
velocities $\dot{\xb}$. In fact, the fluctuations of the thermodynamic 
parameters are governed by the phenomenological Langevin equation (see, e.g., 
in \cite{LanLif}):
$$
\dot x_t = -\l(x_t-\xb) 
           +\sqrt{2\g}~w_t 
$$
which can be generalized for time-dependent $\xb_t$. To lowest order in 
$\dot{\xb}$ the result (\ref{xtret}) comes out. In higher orders the Langevin 
equation gives different results from the present proposal. The standard
distribution functional is the Onsager-Machlup functional \cite{OnsMac53}. 
%
\bibitem{OnsMac53} L. Onsager and S. Machlup, Phys.Rev. {\bf 91}, 1505 (1953); 
{\bf 91}, 1512 (1953); R.L. Stratonovitch, Sel.Transl.Math.Stat.Prob. 
{\bf 10}, 273 (1971); R. Graham, Z.Phys. {\bf B26}, 281 (1976).
%
\bibitem{FT} D.J.Evans, E.G.D.Cohen and G.P.Morriss,
Phys.Rev.Lett. {\bf 71}, 2401 (1993); G. Gallavotti and E.G.D.Cohen,
Phys.Rev.Lett. {\bf 74}, 2694 (1995); C. Maes, F. Redig and A.Van Moffaert,
J.Math.Phys. {\bf 41}, 1528 (2000) 
%

\end{thebibliography}
\end{document}